\documentclass[aps,preprint,tightenlines,superscriptaddress,showpacs]{revtex4}
\usepackage{times}
\usepackage{amsmath}
\usepackage{graphicx}
\usepackage{epsfig}
\usepackage{dcolumn}
\usepackage{bm}
\newcommand{\BR}{{\cal B}}
\newcommand{\pp}{\pi^+\pi^-}

\newcommand{\kk}{K^+K^-}
\newcommand{\LL}{\ell^+\ell^-}
\newcommand{\EE}{e^+e^-}
\newcommand{\MM}{\mu^+\mu^-}

\newcommand{\psip}{\psi(2S)}

\newcommand{\jpsi}{J/\psi}
\newcommand{\ppjpsi}{\pi^+\pi^-J/\psi}
\newcommand{\y}{Y(4260)}
\newcommand{\z}{Z(3900)^\pm}

\def\Journal#1#2#3#4{{#1} {\bf #2}, #3 (#4)}

\def\PRL{Phys. Rev. Lett.}
\def\PRD{Phys. Rev. D}

\def\EPJC{Eur. Phys. J. C}

\begin{document}


\title{\quad\\[0.0cm]
\boldmath
Study of $e^+ e^- \to \pi^+ \pi^- J/\psi$ and  Observation of a Charged Charmonium-like State at Belle}


\noaffiliation
\affiliation{University of the Basque Country UPV/EHU, 48080 Bilbao}
\affiliation{University of Bonn, 53115 Bonn}
\affiliation{Budker Institute of Nuclear Physics SB RAS and Novosibirsk State University, Novosibirsk 630090}
\affiliation{Faculty of Mathematics and Physics, Charles University, 121 16 Prague}
\affiliation{University of Cincinnati, Cincinnati, Ohio 45221}
\affiliation{Deutsches Elektronen--Synchrotron, 22607 Hamburg}
\affiliation{Justus-Liebig-Universit\"at Gie\ss{}en, 35392 Gie\ss{}en}
\affiliation{Gifu University, Gifu 501-1193}
\affiliation{II. Physikalisches Institut, Georg-August-Universit\"at G\"ottingen, 37073 G\"ottingen}
\affiliation{Gyeongsang National University, Chinju 660-701}
\affiliation{Hanyang University, Seoul 133-791}
\affiliation{University of Hawaii, Honolulu, Hawaii 96822}
\affiliation{High Energy Accelerator Research Organization (KEK), Tsukuba 305-0801}
\affiliation{Ikerbasque, 48011 Bilbao}
\affiliation{Indian Institute of Technology Guwahati, Assam 781039}
\affiliation{Indian Institute of Technology Madras, Chennai 600036}
\affiliation{Institute of High Energy Physics, Chinese Academy of Sciences, Beijing 100049}
\affiliation{Institute of High Energy Physics, Vienna 1050}
\affiliation{Institute for High Energy Physics, Protvino 142281}
\affiliation{INFN - Sezione di Torino, 10125 Torino}
\affiliation{Institute for Theoretical and Experimental Physics, Moscow 117218}
\affiliation{J. Stefan Institute, 1000 Ljubljana}
\affiliation{Kanagawa University, Yokohama 221-8686}
\affiliation{Institut f\"ur Experimentelle Kernphysik, Karlsruher Institut f\"ur Technologie, 76131 Karlsruhe}
\affiliation{Kennesaw State University, Kennesaw, Georgia 30144}
\affiliation{Korea Institute of Science and Technology Information, Daejeon 305-806}
\affiliation{Korea University, Seoul 136-713}
\affiliation{Kyungpook National University, Daegu 702-701}
\affiliation{\'Ecole Polytechnique F\'ed\'erale de Lausanne (EPFL), Lausanne 1015}
\affiliation{Faculty of Mathematics and Physics, University of Ljubljana, 1000 Ljubljana}
\affiliation{University of Maribor, 2000 Maribor}
\affiliation{Max-Planck-Institut f\"ur Physik, 80805 M\"unchen}
\affiliation{School of Physics, University of Melbourne, Victoria 3010}
\affiliation{Moscow Physical Engineering Institute, Moscow 115409}
\affiliation{Moscow Institute of Physics and Technology, Moscow Region 141700}
\affiliation{Graduate School of Science, Nagoya University, Nagoya 464-8602}
\affiliation{Kobayashi-Maskawa Institute, Nagoya University, Nagoya 464-8602}
\affiliation{Nara Women's University, Nara 630-8506}
\affiliation{National Central University, Chung-li 32054}
\affiliation{National United University, Miao Li 36003}
\affiliation{Department of Physics, National Taiwan University, Taipei 10617}
\affiliation{H. Niewodniczanski Institute of Nuclear Physics, Krakow 31-342}
\affiliation{Nippon Dental University, Niigata 951-8580}
\affiliation{Niigata University, Niigata 950-2181}
\affiliation{Osaka City University, Osaka 558-8585}
\affiliation{Pacific Northwest National Laboratory, Richland, Washington 99352}
\affiliation{Panjab University, Chandigarh 160014}
\affiliation{University of Pittsburgh, Pittsburgh, Pennsylvania 15260}
\affiliation{RIKEN BNL Research Center, Upton, New York 11973}
\affiliation{University of Science and Technology of China, Hefei 230026}
\affiliation{Seoul National University, Seoul 151-742}
\affiliation{Sungkyunkwan University, Suwon 440-746}
\affiliation{School of Physics, University of Sydney, NSW 2006}
\affiliation{Tata Institute of Fundamental Research, Mumbai 400005}
\affiliation{Excellence Cluster Universe, Technische Universit\"at M\"unchen, 85748 Garching}
\affiliation{Toho University, Funabashi 274-8510}
\affiliation{Tohoku Gakuin University, Tagajo 985-8537}
\affiliation{Tohoku University, Sendai 980-8578}
\affiliation{Department of Physics, University of Tokyo, Tokyo 113-0033}
\affiliation{Tokyo Institute of Technology, Tokyo 152-8550}
\affiliation{Tokyo Metropolitan University, Tokyo 192-0397}
\affiliation{Tokyo University of Agriculture and Technology, Tokyo 184-8588}
\affiliation{University of Torino, 10124 Torino}
\affiliation{CNP, Virginia Polytechnic Institute and State University, Blacksburg, Virginia 24061}
\affiliation{Wayne State University, Detroit, Michigan 48202}
\affiliation{Yamagata University, Yamagata 990-8560}
\affiliation{Yonsei University, Seoul 120-749}
  \author{Z.~Q.~Liu}\affiliation{Institute of High Energy Physics, Chinese Academy of Sciences, Beijing 100049} 
  \author{C.~P.~Shen}\altaffiliation[now at ]{Beihang University, Beijing 100191}{\affiliation{Graduate School of Science, Nagoya University, Nagoya 464-8602}
  \author{C.~Z.~Yuan}\affiliation{Institute of High Energy Physics, Chinese Academy of Sciences, Beijing 100049} 
  \author{I.~Adachi}\affiliation{High Energy Accelerator Research Organization (KEK), Tsukuba 305-0801} 
  \author{H.~Aihara}\affiliation{Department of Physics, University of Tokyo, Tokyo 113-0033} 
  \author{D.~M.~Asner}\affiliation{Pacific Northwest National Laboratory, Richland, Washington 99352} 
  \author{V.~Aulchenko}\affiliation{Budker Institute of Nuclear Physics SB RAS and Novosibirsk State University, Novosibirsk 630090} 
  \author{T.~Aushev}\affiliation{Institute for Theoretical and Experimental Physics, Moscow 117218} 
  \author{T.~Aziz}\affiliation{Tata Institute of Fundamental Research, Mumbai 400005} 
  \author{A.~M.~Bakich}\affiliation{School of Physics, University of Sydney, NSW 2006} 
  \author{A.~Bala}\affiliation{Panjab University, Chandigarh 160014} 
  \author{K.~Belous}\affiliation{Institute for High Energy Physics, Protvino 142281} 
% ...
\collaboration{The Belle Collaboration}

\begin{abstract}

The cross section for $e^+ e^- \to \pi^+ \pi^- J/\psi$ between 3.8~GeV and 5.5~GeV is
measured with a 967~fb$^{-1}$ data sample collected by the Belle detector at or near the
$\Upsilon(nS)$ ($n = 1,\ 2,\ ...,\ 5$) resonances.  The $Y(4260)$ state is observed, and its
resonance parameters are determined. In addition, an excess of
$\pi^+ \pi^- J/\psi$ production around 4~GeV is observed.
This feature  can be described by a Breit-Wigner
parameterization  with properties that are consistent
with the $Y(4008)$ state that was previously reported by Belle.
In a study of $Y(4260) \to \pi^+ \pi^- J/\psi$
decays, a structure is observed in the
$M(\pi^\pm\jpsi)$ mass spectrum with $5.2\sigma$ significance, with mass $M=(3894.5\pm 6.6\pm 4.5)~{\rm
MeV}/c^2$ and width $\Gamma=(63\pm 24\pm 26)$~MeV/$c^{2}$, where the
errors are statistical and systematic, respectively.
This structure can be interpreted as a new
charged charmonium-like state.

\end{abstract}


\maketitle


The $\y$ state was first observed by the BaBar
Collaboration in the initial-state-radiation (ISR) process $\EE \to
\gamma_{\rm ISR}\ppjpsi$~\cite{babay4260} and then
confirmed by the CLEO~\cite{cleo_y} and Belle
experiments~\cite{belle_y} using the same technique.
Subsequently, a charged $Z(4430)^\pm$ charmonium-like state was reported in the
$\pi^\pm\psip$ invariant mass spectrum  of $B \to K\pi^{\pm} \psip$~\cite{belle_z4430}
and two $Z^{\pm}$ states were observed in the $\pi^{\pm} \chi_{c1}$
invariant mass distribution  of $B \to K \chi_{c1} \pi^{\pm}$~\cite{twoz}.
Motivated by the striking
observations of charged charmonium-like~\cite{belle_z4430, twoz}
and bottomonium-like states~\cite{zb}, we
investigate the existence of
similar states as intermediate
resonances in $\y\to \ppjpsi$ decays.

After the initial observations of the $Y(4260)$~\cite{babay4260, cleo_y, belle_y},
CLEO collected 13.2~pb$^{-1}$ of $e^+ e^-$ data at
$\sqrt{s}=4.26$~GeV and investigated 16 possible $Y(4260)$  decay modes with
charmonium or light hadrons in the final state~\cite{cleoy4260}. An ISR analysis by the Belle experiment with
548~fb$^{-1}$ of data collected at or near $\sqrt{s}=10.58$~GeV~\cite{belley4260}
showed a significant $\y$ signal as well as an excess of $\ppjpsi$
event production near 4~GeV that could be
described by a broad Breit-Wigner (BW) parameterization --- the so-called
$Y(4008)$. Recently, the BaBar Collaboration
reported an updated ISR analysis  with 454~fb$^{-1}$ of data and a modified
approach for the background description~\cite{babarnew}; the $\y$
state was observed with improved significance, but the $Y(4008)$
structure was not confirmed. Instead, they
attributed the structure below the $Y(4260)$ to
exponentially falling
non-resonant $\ppjpsi$ production.

In this Letter, we report cross section measurements for $\EE\to
\ppjpsi$ between 3.8~GeV and 5.5~GeV, and a search for structures in
the $\ppjpsi$, $\pi^{\pm}\jpsi$, and $\pp$ systems. The
results are based on the full Belle data sample with an integrated
luminosity of 967~fb$^{-1}$ collected at or near the $\Upsilon(nS)$ resonances
($n={1, 2, ..., 5}$). The Belle detector operated at the KEKB
asymmetric-energy $\EE$ collider~\cite{kekb} and is described in
detail elsewhere~\cite{belle-detector}. We use the {\sc phokhara}~\cite{phokhara}
program to generate signal Monte Carlo
(MC) events and determine experimental efficiencies.
The results reported here supersede those of Ref.~\cite{belley4260}, wherein
a subset of the Belle data sample was used.


The event selection is described in
Ref.~\cite{belley4260}. We require
four well reconstructed charged tracks with zero net charge.
For each charged track,
a likelihood ${\mathcal{L}}_X$ is formed from different detector subsystems
for particle hypothesis $X \in \{ e,\ \mu,\ \pi,\ K,\ p \}$.
Tracks with a likelihood ratio $\mathcal{R}_K = \frac{\mathcal{L}_K}
{\mathcal{L}_K+\mathcal{L}_\pi} < 0.4$ are identified as pions
with an efficiency of about 95\%. Similar ratios are also defined
for lepton-pion discrimination~\cite{EIDMUID}. For electrons from $\jpsi\to \EE$, one
track should have $\mathcal{R}_e>0.95$ and the other track
$\mathcal{R}_e>0.05$. For muons from $\jpsi\to \MM$, at least one
track should have $\mathcal{R}_\mu>0.95$; in cases where the other track
has no muon identification, in order to suppress misidentified  muon tracks,
the polar angles of the two muon tracks in the $\pp \MM$
center-of-mass (CM) frame must satisfy $|\cos\theta_{\mu}|<0.7$. Events
with $\gamma$ conversions are removed by requiring
$\mathcal{R}_e<0.75$ for the $\pp$ candidate tracks. Furthermore, in $\jpsi\to \EE$,
such events are further reduced by requiring the
invariant mass of the $\pp$ candidate pair to be larger than 0.32~GeV/$c^2$. Events with
a total energy deposit in the electromagnetic calorimeter (ECL)
above 9~GeV are removed in the $\jpsi\to \EE$ mode
because the MC simulation of
the trigger efficiency for these
Bhabha-like events does not accurately reproduce the data.
There is only one combination of $\pp\LL$ ($\ell=e,~\mu$) in each
event after the above selections.

Candidate  ISR events are identified by the requirement
$|M^2_{\rm rec}|<2.0$~(GeV/$c^2$)$^2$, where $M^2_{\rm rec} =
(P_{CM}-P_{\pi^+}-P_{\pi^-}-P_{\ell^+}-P_{\ell^-})^2$ and $P_i$
represents the four-momentum of the corresponding particle or composite in the
$\EE$ CM frame. Clear $\jpsi$ signals are observed in both
the $\jpsi\to \EE$ and $\MM$ modes. We define the $\jpsi$
signal region as $3.06~\hbox{GeV}/c^2 <M(\LL)<3.14$~GeV/$c^2$ (the mass resolution for
lepton pairs being about 20~MeV/$c^2$), and $\jpsi$ mass sidebands as
$2.91 ~\hbox{GeV}/c^2<M(\LL)<3.03$~GeV/$c^2$ or $3.17~\hbox{GeV}/c^2<M(\LL)<3.29$~GeV/$c^2$,
which are three times as wide as the signal region.

Figure~\ref{mppjpsi}(a) shows the $\pp\LL$ invariant
mass~\cite{def-mass} distributions after
all of these selection requirements are applied.
Also shown in this figure are the background estimates evaluated using
the normalized $\jpsi$-mass sidebands. Two enhancements --- the $Y(4008)$ and the
$Y(4260)$ --- above 3.8~GeV/$c^2$ are observed,
consistent with the results of
Ref.~\cite{belley4260} but in disagreement with those of Ref.~\cite{babarnew}.
Other possible background sources not included in the
sidebands are found to be small from MC simulation~\cite{cleoy4260}; these include
(1) $\ppjpsi$ with $\jpsi$ decays into final
states other than lepton pairs and (2) $X\jpsi$, with $X$ not being
a $\pp$ pair, such as $\kk$ or $\pp\pi^0$. Non-ISR
production of $\EE\to \ppjpsi$ final states, such as $\EE\to
\gamma\gamma^*\gamma^*\to \gamma\rho^0\jpsi$, is also estimated to be
small~\cite{zhuk}.
Figure~\ref{mppjpsi}(b) shows the measured cross
sections for $\EE \to \ppjpsi$, where the error bars are statistical only.

\begin{figure*}[htbp]
\begin{center}
\includegraphics[width=0.43\textwidth]{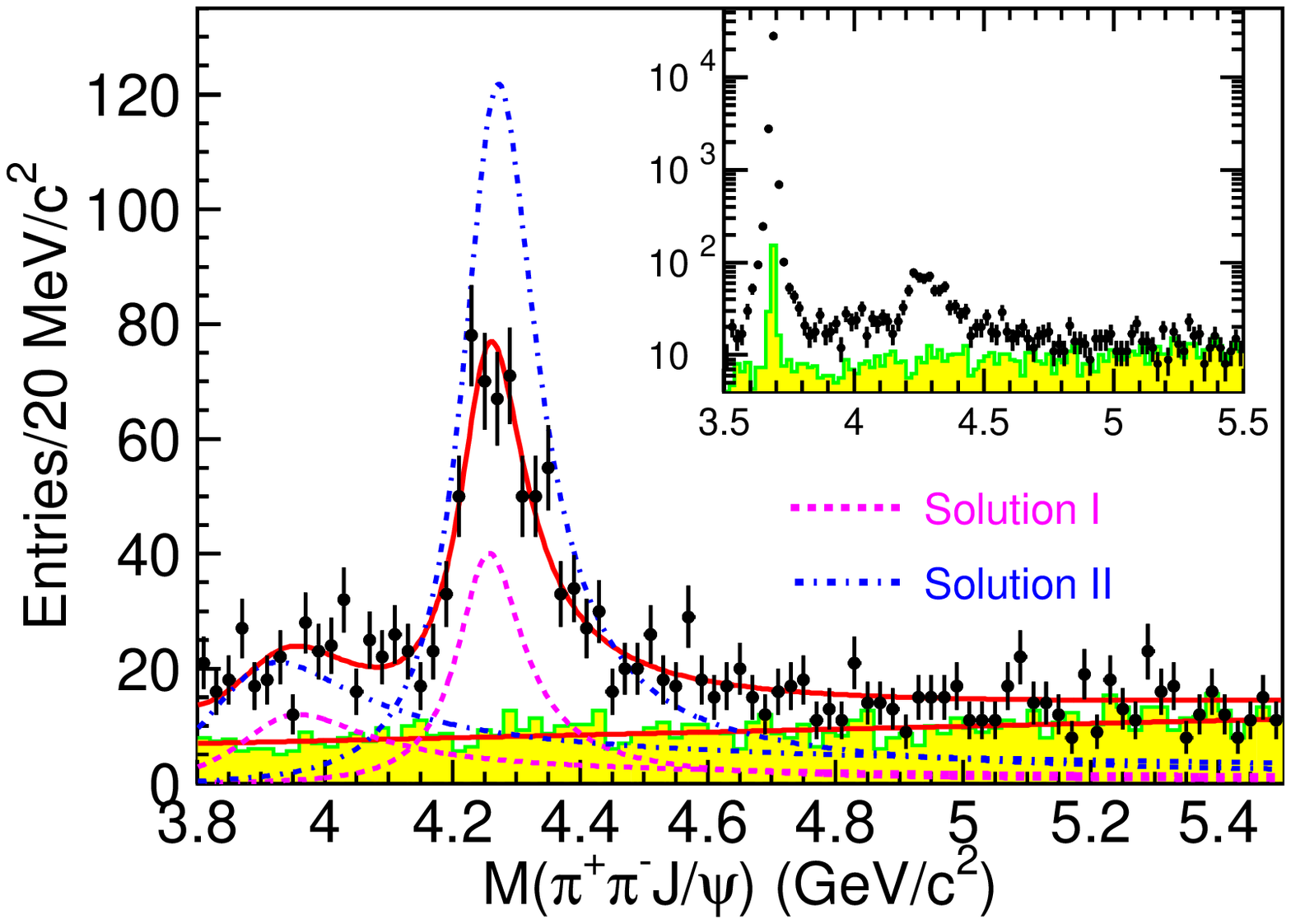}
\includegraphics[width=0.45\textwidth]{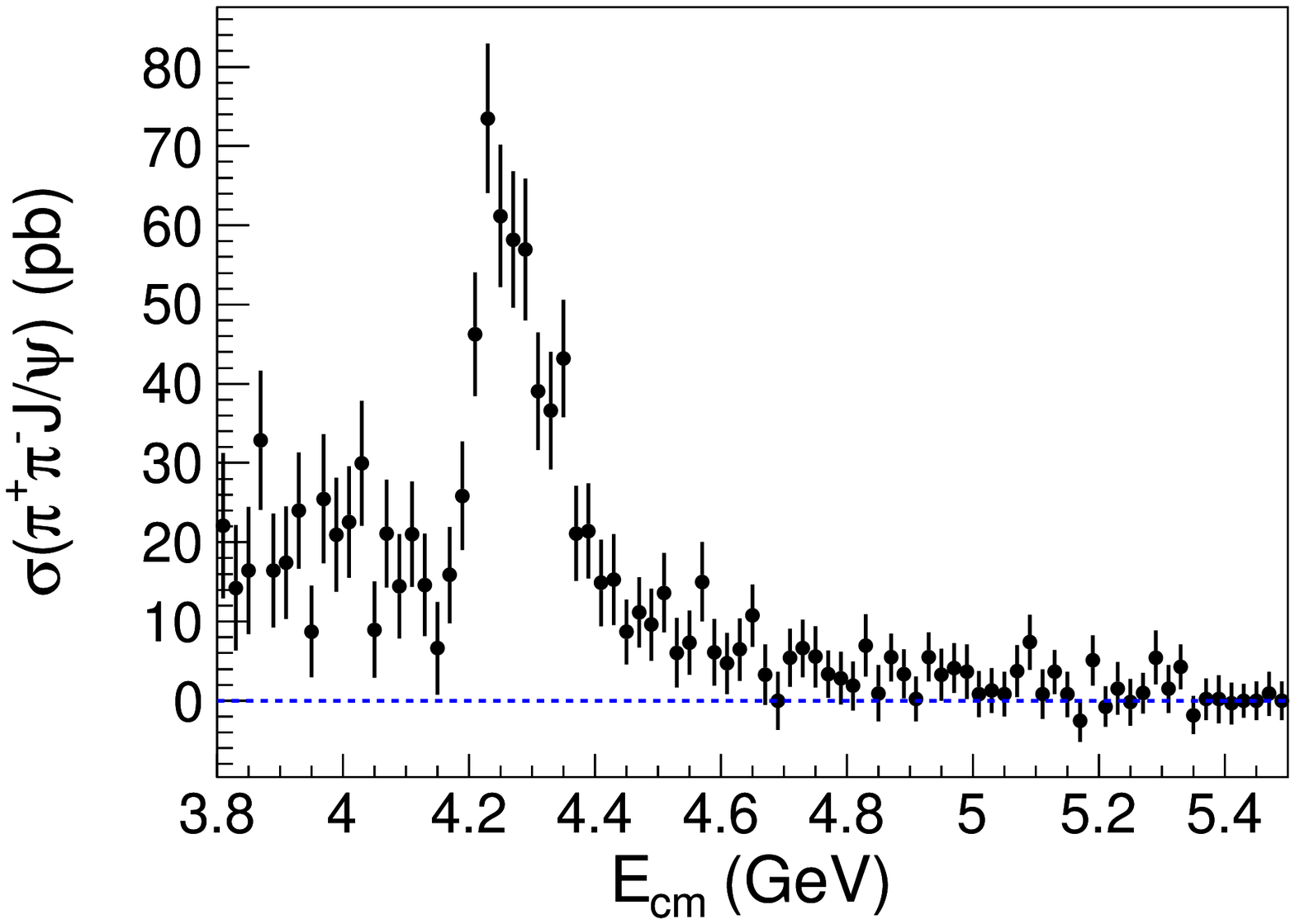}
  \put(-370,120){\bf (a)}
  \put(-60,120){\bf (b)}
\caption{(a) Invariant mass distributions of $\pp\LL$. Points with error bars
are data, and the shaded histograms are the normalized $\jpsi$
mass sidebands. The solid curves show
the total best fit with two coherent
resonances and contribution from background.
The dashed curves are for solution I, while the dot-dashed curves are for
solution II.  The inset shows the distributions on a
logarithmic vertical scale.
The large peak around 3.686~GeV/$c^2$ is the $\psip\to
\ppjpsi$ signal. (b) Cross section  of $\EE\to \ppjpsi$ after background subtraction.
The errors are statistical only.} \label{mppjpsi}
\end{center}
\end{figure*}

Systematic uncertainties of the cross
section measurement are
found to be 7.9\% and 7.3\% for the $\EE$ and $\mu^+ \mu^-$ modes, respectively.
The particle identification (PID) uncertainties, measured from pure
$\psip$ events in the same data sample, are 4.7\% and 3.6\% for
the $\EE$ and $\MM$ modes, respectively. Tracking efficiency
uncertainties are estimated to be 3.3\% for both $\EE$ and $\MM$
modes in the momentum and angular regions of interest for signal
events. The uncertainties associated with the choice of the $\jpsi$ mass window and
$|M^2_{\rm rec}|$ requirements are also estimated using pure $\psip$
events. It is found that MC efficiencies are higher than data by
($4.5\pm 0.4$)\% in the $\EE$ mode and $(4.1\pm 0.2)$\% in the
$\MM$ mode. The differences in efficiencies are corrected and the
uncertainties in the correction factors are incorporated into the systematic
errors. Overall, together with the $|M^2_{\rm rec}|$ requirements, these uncertainties
contribute 0.6\% for the $\EE$ mode and 0.3\% for
the $\MM$ mode within the $\jpsi$ mass window. Belle measures luminosity with a
precision of 1.4\% using wide-angle Bhabha events. The {\sc
phokhara} generator calculates ISR with 0.1\% accuracy~\cite{phokhara}.
The dominant uncertainties due to the MC
generator are from three-body decay dynamics.  MC simulation
with modified $\pp$ invariant mass distributions weighted according to data
distributions yields a 2\% to 5\% efficiency difference compared
with a phase space $\pp$ mass spectrum model. Thus, we
conservatively use 5\% as the systematic error due to the
approximations made in the MC event
generator. According to MC simulation, the offline trigger efficiency
for four-track events of the studied topology exceeds 99\%.
A 1.0\% systematic error is included for the trigger uncertainty. The
uncertainty of $\BR(\jpsi\to \LL)=\BR(\jpsi\to \EE)+ \BR(\jpsi\to
\MM)$ is 1.0\%~\cite{PDG}.

As a validation of our analysis, we also measure the ISR $\psip$
production rate using the same selection criteria. The cross
sections are $(14.12\pm 0.18\pm 0.85)$~pb and $(15.13\pm 0.11\pm
0.79)$~pb at $\sqrt{s}=10.58$~GeV~\cite{cs} for the $\EE$ and $\MM$ modes, respectively.
Our measurement agrees within errors with the prediction of $(14.25\pm 0.26)$~pb~\cite{kuraev}
using the world-average resonance parameters~\cite{PDG}.

An unbinned maximum likelihood fit is performed to the $\ppjpsi$
mass spectrum above 3.8~GeV/$c^2$. As there are two enhancements
observed, as shown in Fig.~\ref{mppjpsi}(a), we use the same fit strategy as in
Ref.~\cite{belley4260}. Two coherent BW functions ($R_1$, $R_2$) are used to
describe the $Y(4008)$ and $Y(4260)$ structures, assuming there is no continuum production
of $\EE\to\ppjpsi$. In the fit, the background term is fixed at the level obtained from a
linear fit to the sideband data.
The solid curves in Fig.~\ref{mppjpsi}(a) show the fit results. There are two
solutions of equal optimum fit quality. The masses and widths of the
resonances are the same for the two solutions; the partial widths
to $\EE$ and the relative phase between the two resonances are
different (see Table~\ref{par-res})~\cite{BeeBr}.  The
fit quality is estimated using the reduced $\chi^2$ statistic; we obtain
$\chi^2/ndf=101.6/84$, corresponding to a confidence level
of 9.3\%. Systematic uncertainties in the extracted values of the resonance parameters arise
from the absolute energy scale, the detection
efficiency, background estimation and the parameterization of the
resonance models. The absolute energy uncertainty is estimated
from the $\psip$ mass fit.
Uncertainty in the detection efficiency does not affect the mass and
width measurements, but could affect the measurement of the partial width to $\EE$.
Systematic uncertainties associated with the background contribution are estimated
by varying the background level by $\pm 1\sigma$ in the fit.
Resonance parameterization is studied
by changing the $Y(4260)$ BW function from a parameterization
with a constant width to another with a three-body
phase-space-dependent function.
The interference between the two resonances,
$Y(4260)$ and $Y(4008)$, depends on the structure of the $\ppjpsi$ amplitude,
which can be different for the two resonances. We conservatively estimate
possible systematic effects by performing a fit without the interference
between the $Y(4008)$ and the $Y(4260)$ and taking the difference compared to
results with interference as the corresponding systematic error.
All of these contributions are summarized in Table~\ref{par-res}.
The measured mass, width and the product $\Gamma_{ee}\mathcal{B}(R_1\to\ppjpsi)$
are consistent within statistics with the previous results~\cite{belley4260, small-err}.

\begin{table}[htbp]
\caption{Results of the fits to the $\ppjpsi$ mass spectrum with two
coherent resonances. $M(R_i)$, $\Gamma_{{\mathrm{tot}}}(R_i)$ and
$\Gamma_{ee} \BR(R_i\to \ppjpsi)$, $i=1,2$ represent the mass
(in MeV/$c^2$), total width (in MeV/$c^2$) and product of the
branching ratio for the decay into $\ppjpsi$ and the $\EE$ partial width (in
eV/$c^2$) for the two resonances, respectively. The parameter $\phi$ (in degrees)
is the relative phase between the two resonances. The first and second errors
are statistical and systematic, respectively.} \label{par-res}
\begin{tabular}{c c c}
\hline\hline
Parameters & Solution I & Solution II \\
\hline
$M(R_1)$ & \multicolumn{2}{c}{$3890.8\pm40.5\pm 11.5$} \\
$\Gamma_{{\mathrm{tot}}}(R_1)$ & \multicolumn{2}{c}{$254.5\pm39.5\pm 13.6$} \\
$\Gamma_{ee}\mathcal{B}(R_1\to\ppjpsi)$ & ($3.8\pm0.6\pm0.4$) & ($8.4\pm1.2\pm1.1$) \\[6pt]
$M(R_2)$ & \multicolumn{2}{c}{$4258.6\pm8.3\pm12.1$} \\
$\Gamma_{{\mathrm{tot}}}(R_2)$ & \multicolumn{2}{c}{$134.1\pm16.4\pm5.5$} \\
$\Gamma_{ee}\mathcal{B}(R_2\to\ppjpsi)$ & ($6.4\pm0.8\pm0.6$) & ($20.5\pm1.4\pm2.0$) \\[6pt]
$\phi$ & $59\pm17\pm11$ & $-116\pm6\pm11$ \\
\hline\hline
\end{tabular}
\end{table}

Figure~\ref{dalitz} shows the Dalitz plot for events in the $\y$
signal region ($4.15~\hbox{GeV}/c^2  < M(\ppjpsi) < 4.45$~GeV/$c^2$), where
we observe structures in the $\pp$ and $\pi^+\jpsi$ systems. The inset is
for the events in the $\jpsi$-mass sidebands, where no obvious
structures are observed in the non-$\ppjpsi$ background events.

\begin{figure}[htbp]
\includegraphics[width=0.45\textwidth]{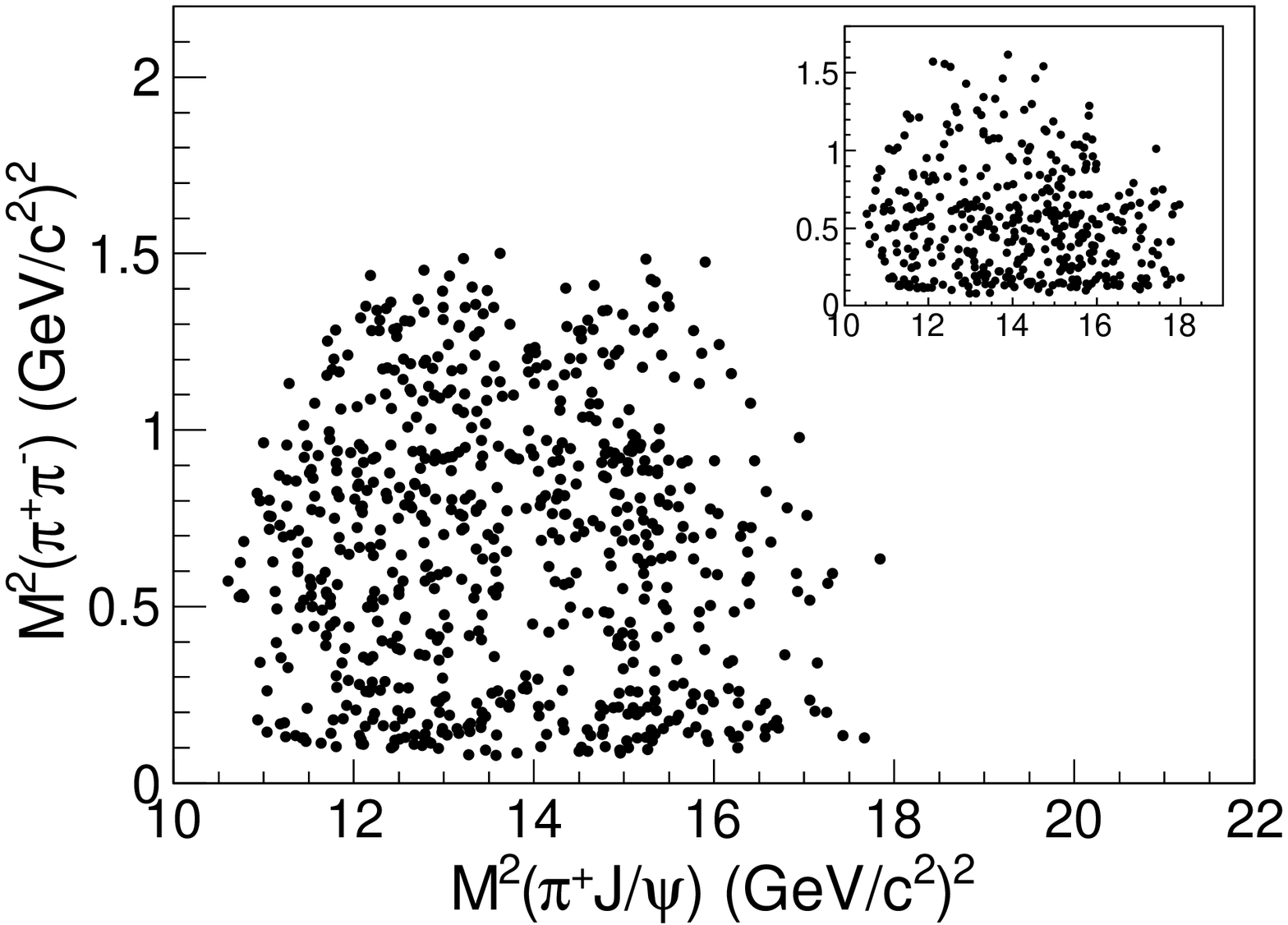}
\caption{Dalitz plot for $\y\to \ppjpsi$ decays for $4.15~\hbox{GeV}/c^2 <
M(\ppjpsi) < 4.45$~GeV/$c^2$. The inset shows background events
from the $\jpsi$-mass sidebands (not normalized). } \label{dalitz}
\end{figure}

Figure~\ref{proj} shows a projection of the $M(\pp)$,
$M(\pi^+\jpsi)$ and $M(\pi^-\jpsi)$ invariant mass distributions
for events in the $\y$ signal region. Background contributions are estimated
from the normalized $\jpsi$ mass sidebands. There are $f_0(980)$,
$f_0(500)$ and non-resonant S-wave amplitudes in the $\pp$ mass
spectrum. In the $\pi^\pm\jpsi$ mass spectrum, there is a
significant peak around 3.9~GeV/$c^2$ (called the $\z$ hereafter)
that we interpret as evidence for an exotic charmonium-like state decays into $\pi^\pm\jpsi$; the
broader peak near 3.5~GeV/$c^2$ is a reflection of the $\z$,
as confirmed by MC simulation.
The pure $\pp$ S-wave amplitudes that describe the $\pp$ invariant mass
distribution well cannot reproduce the structure at 3.9~GeV/$c^2$ in the $\pi^{\pm}\jpsi$ mass spectra,
as shown in Fig.~\ref{proj} with the open histograms.

\begin{figure*}[htbp]
 \includegraphics[width=0.32\textwidth]{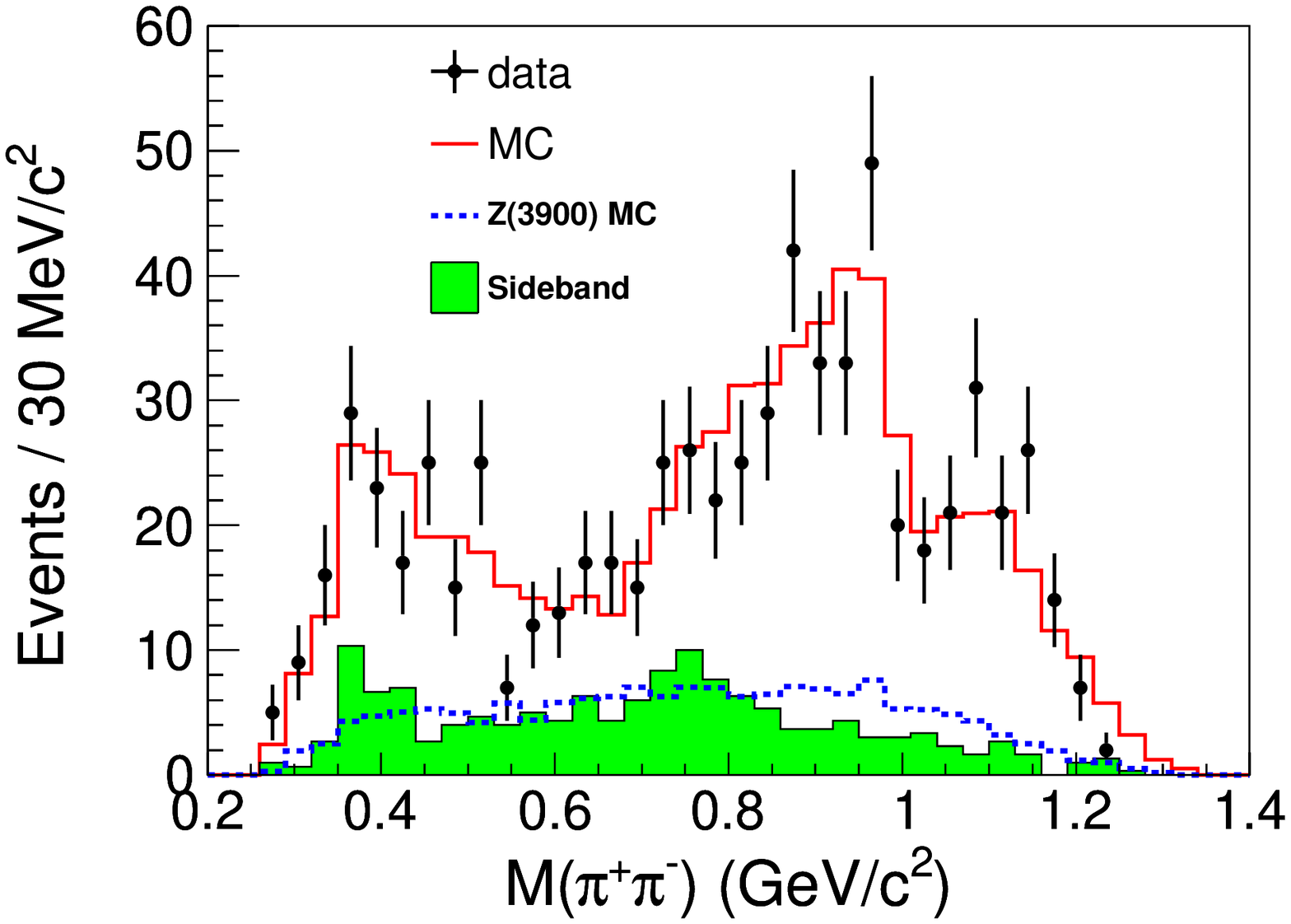}
 \put(-30,90){ \bf (a)}
 \includegraphics[width=0.32\textwidth]{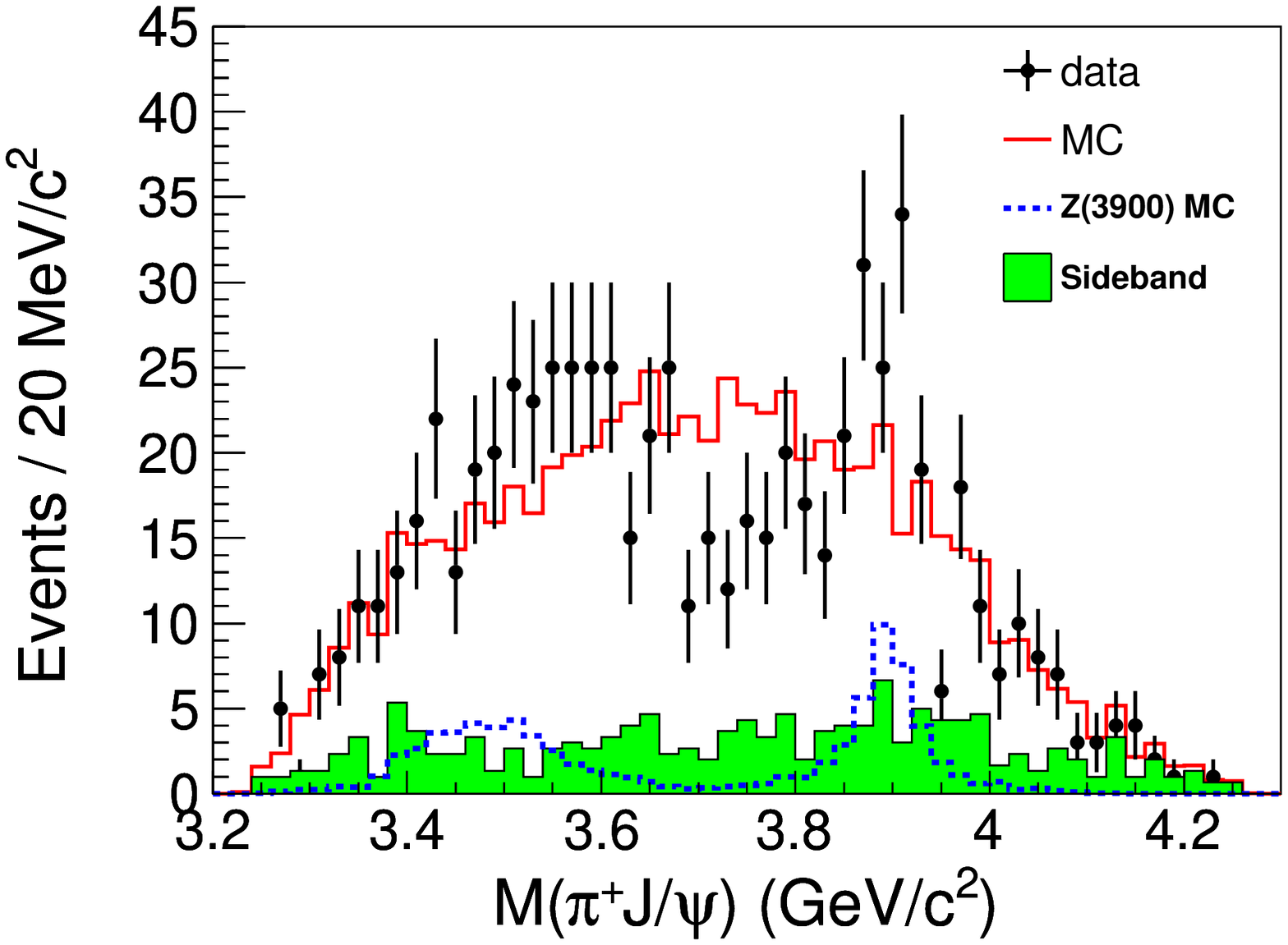}
 \put(-120,90){ \bf (b)}
 \includegraphics[width=0.32\textwidth]{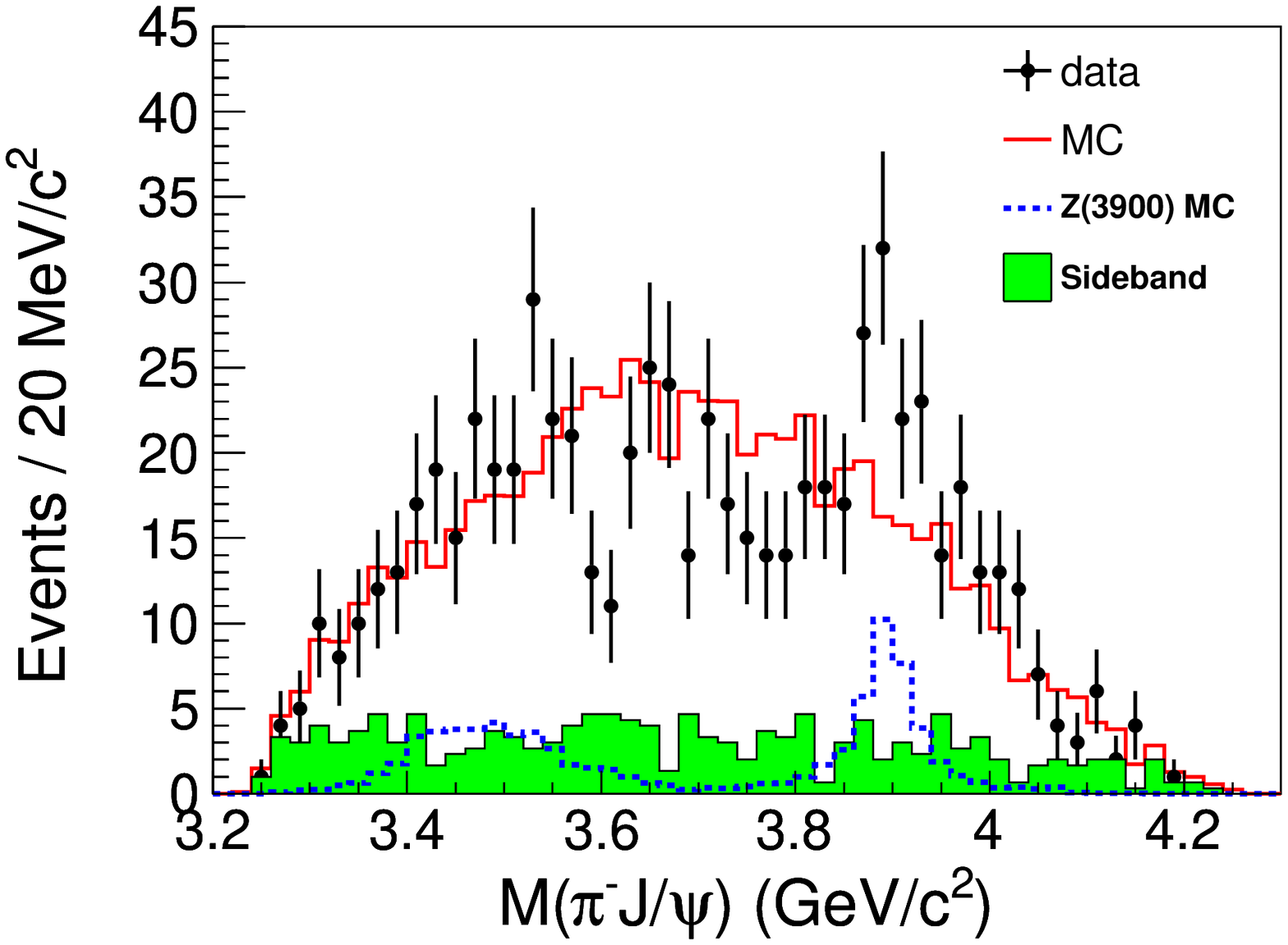}
  \put(-120,90){ \bf (c)}
\caption{Invariant mass distributions of (a) $\pp$, (b) $\pi^+\jpsi$ and (c)
$\pi^-\jpsi$ for events in the $\y$ signal region. Points with error
bars represent data, shaded histograms are normalized background estimates from
the $\jpsi$-mass sidebands, solid histograms represent
MC simulations of $\pp$  amplitudes~\cite{pwa} (normalized $\jpsi$-mass sideband
events added) and dashed histograms
are MC simulation results for a $\z$ signal}. \label{proj}
\end{figure*}

An unbinned maximum likelihood fit is performed to the
distribution of $M_{\mathrm{max}}(\pi J/\psi)$,  the maximum of
$M(\pi^+\jpsi)$ and $M(\pi^-\jpsi)$. The signal shape
is parameterized as an S-wave BW function convolved with a
Gaussian whose mass resolution is fixed at the MC-estimated value of
7.4~MeV, and the background is approximated by a cubic polynomial. The mass-dependent
efficiency is also included in the fit.
Figure~\ref{projfit} shows the fit results. The fit yields $159\pm
49 \pm 7$ $\z$ events, with a mass of $(3894.5\pm 6.6\pm 4.5)~{\rm
MeV}/c^2$ and a width of $(63\pm 24\pm 26)$~MeV/$c^2$, where the
errors are statistical and systematic, respectively. The
largest contributions to the systematic uncertainties arise from the parameterization of the
signal and background shapes, the mass resolution and mass
calibrations. The statistical
significance of the $\z$ state is found to be $5.5\sigma$ in the
nominal fit, and is larger than $5.2\sigma$
in all alternate fits used for
systematic checks with different background shapes, fit ranges and BW resonant models.
The significance is calculated by comparing the
logarithmic likelihoods with and without the $\z$ signal,
and including the change of the number of parameters in the fits.
From the signal yields and the MC-simulated efficiencies, we
obtain the ratio of the production rates $\frac{\BR(\y\to
\z\pi^\mp) \BR(\z\to \pi^\pm \jpsi)}{\BR(\y\to \ppjpsi)}=(29.0\pm 8.9)\%$,
where the error is statistical
only~\cite{ratio}. We test the hypothesis that interference
between the S- and D-waves in the $\pp$ system might produce a
structure similar to the enhancement observed in data. Although the statistics
do not allow us to fully explore the Dalitz plot via an amplitude analysis,
we find that these partial waves alone cannot produce a $\pi^{\pm}\jpsi$ invariant mass peak near 3.9~GeV/$c^2$~\cite{pwa}.
Inclusion of a $\z$ with the mass and width determined above
significantly improves the agreement between predicted and observed Dalitz plot distributions.

\begin{figure}[htbp]
 \includegraphics[height=5.3cm]{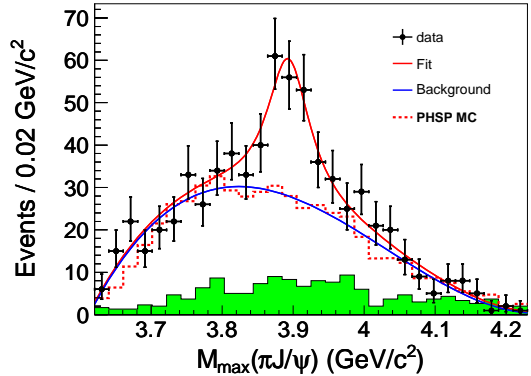}
\caption{Unbinned maximum likelihood fit to the
distribution of the $M_{\mathrm{max}}(\pi J/\psi)$. Points with error
bars are data, the curves are the best fit, the dashed histogram
is the phase space distribution and the shaded histogram is the
non-$\ppjpsi$ background estimated from the normalized $\jpsi$
sidebands.} \label{projfit}
\end{figure}

In summary, the cross section of $\EE\to \ppjpsi$ is measured from
3.8~GeV to 5.5~GeV. The $\y$ resonance is observed
and its resonant parameters are determined. In addition, the
$Y(4008)$ state is confirmed. The intermediate states in
$\y\to \ppjpsi$ decays are also investigated. A $\z$ state with
a mass of $(3894.5\pm 6.6\pm 4.5)~{\rm MeV}/c^2$ and a width of $(63\pm 24\pm
26)$~MeV/$c^2$ is observed in the $\pi^\pm\jpsi$ mass spectrum with a
statistical significance larger than $5.2\sigma$. This state is
close to the $D\bar{D}^*$ mass threshold; however, no enhancement is observed near the
$D^*\bar{D}^*$ mass threshold.
As the $\z$ state has a strong coupling to charmonium and is charged,
we conclude it cannot be a conventional $c\bar{c}$ state.


We thank the KEKB group for excellent operation of the
accelerator; the KEK cryogenics group for efficient solenoid
operations; and the KEK computer group, the NII, and PNNL/EMSL for
valuable computing and SINET4 network support. We acknowledge
support from MEXT, JSPS and Nagoya's TLPRC (Japan); ARC and DIISR
(Australia); NSFC (China); MSMT (Czechia); DST (India); INFN
(Italy); MEST, NRF, GSDC of KISTI, and WCU (Korea); MNiSW and NCN
(Poland); MES and RFAAE (Russia); ARRS (Slovenia); SNSF
(Switzerland); NSC and MOE (Taiwan); and DOE and NSF (USA).
This work is supported partly by a Grant-in-Aid from MEXT for
Science Research on Innovative Areas (``Elucidation of New
Hadrons with a Variety of Flavors") and JSPS KAKENHI Grant
No. 24740158.

{\it Note added}.-- As we were preparing to submit this paper, we became aware of
a paper from the BESIII Collaboration~\cite{bes3zc} that also reports on
the $\z$.



\begin{thebibliography}{**}

\bibitem{babay4260} B.~Aubert {\em et al.} (BaBar Collaboration),
\Journal\PRL{95}{142001}{2005}.

\bibitem{cleo_y} Q.~He {\em et al.} (CLEO Collaboration),
\Journal\PRD{74}{091104(R)}{2006}.

\bibitem{belle_y} K.~Abe {\em et al.} (Belle Collaboration),
hep-ex/0612006.


\bibitem{belle_z4430} S.~K.~Choi {\em et al.} (Belle Collaboration),
\Journal\PRL{100}{142001}{2008}.

\bibitem{twoz} R. Mizuk {\em et al.} (Belle Collaboration),
\Journal\PRD{78}{072004}{2008}.

\bibitem{zb} A.~Bondar {\em et al.} (Belle Collaboration),
\Journal\PRL{108}{122001}{2012}.


\bibitem{cleoy4260} T.~E.~Coan {\em et al.} (CLEO Collaboration),
\Journal\PRL{96}{162003}{2006}.

\bibitem{belley4260} C.~Z.~Yuan {\em et al.} (Belle Collaboration),
\Journal\PRL{99}{182004}{2007}.

\bibitem{babarnew} J.~P.~Lees {\em et al.} (BaBar Collaboration),
\Journal\PRD{86}{051102(R)}{2012}.


\bibitem{kekb} S. Kurokawa and E. Kikutani, Nucl. Instrum. Methods
Phys. Res., Sect. A {\bf 499}, 1 (2003), and other papers
included in this volume; T. Abe {\em et al.}, Prog. Theor. Exp. Phys. (2013) 03A001
and following articles up to 03A011.


\bibitem{belle-detector} A.~Abashian {\em et al.} (Belle Collaboration),
 Nucl. Instrum. Methods Phys. Res., Sect. A {\bf 479}, 117 (2002);
 also see detector section in J. Brodzicka {\em et al.}, Prog. Theor. Exp. Phys. (2012) 04D001.


\bibitem{phokhara} G.~Rodrigo, H.~Czy$\dot{\hbox{z}}$, J.~H.~K$\ddot{\hbox{u}}$hn and M.~Szopa,
\Journal\EPJC{24}{71}{2002}.

\bibitem{EIDMUID} K. Hanagaki {\em et al.}, Nucl. Instrum. Methods
Phys. Res., Sect. A {\bf 485}, 490 (2002);
A. Abashian {\em et al.}, Nucl. Instrum. Methods
Phys. Res., Sect. A {\bf 491}, 69
(2002).

\bibitem{def-mass} In the paper, $M(\pp\LL)-M(\LL)+M(\jpsi)$ is
used instead of the invariant mass of the four final state
particles to improve the mass resolution. Here $M(\jpsi)$ is the
nominal mass of the $\jpsi$.

\bibitem{zhuk}  K.~Zhu, C.~Z.~Yuan and R.~G.~Ping,
  Phys.\ Rev.\ D {\bf 78}, 036004 (2008).

\bibitem{PDG} J. Beringer {\itshape et al.} (Particle Data Group),
\Journal\PRD{86}{010001}{2012}.

\bibitem{cs} The cross sections are measured to be
$(13.79\pm 0.44\pm 0.83)$~pb
and $(13.33\pm 0.25\pm 0.70)$~pb at $\sqrt{s}=10.87$~GeV,
$(16.75\pm 0.85\pm 1.01)$~pb and $(16.63\pm 0.54\pm 0.87)$~pb at
$\sqrt{s}=10.02$~GeV, for the $\EE$ and $\MM$ modes, respectively.
Our measurements agree with the predictions of $(13.42\pm 0.25)$~pb at 10.87~GeV, and $(16.03\pm
0.29)$~pb at 10.02~GeV~\cite{kuraev} within errors.

\bibitem{kuraev} E.~A.~Kuraev and
V.~S.~Fadin, Sov.\ J.\ Nucl.\ Phys.\  {\bf 41}, 466 (1985) [Yad.\
Fiz.\ {\bf 41}, 733 (1985)].


\bibitem{BeeBr} Considering the correlation between
$\Gamma_{ee} \mathcal{B}(R\to\ppjpsi)$ and $\Gamma_{{\textrm{tot}}}$, we
get $\mathcal{B}_{ee}\mathcal{B}(R\to\ppjpsi) =
(1.5\pm0.3\pm0.2)\times 10^{-8}$ and $(4.8\pm0.6\pm0.5)\times
10^{-8}$ for $R_1$ and $R_2$, respectively, for solution I; and
$\mathcal{B}_{ee}\mathcal{B}(R\to\ppjpsi) = (3.3\pm0.7\pm0.5)
\times 10^{-8}$ and $(15.3\pm1.4\pm1.5) \times 10^{-8}$ for $R_1$
and $R_2$, respectively, for solution II, where the first and second
errors are statistical and systematic, respectively.

\bibitem{small-err} A smaller systematic error is obtained because we
no longer use a BW with a phase-space-dependent total width to parameterize $R_1$.

\bibitem{pwa} We perform a partial wave analysis with $f_0(500)$, $f_0(980)$,
non-resonant S-wave, and $f_2(1270)$ amplitudes, and find that the S-wave contributions
dominate. The MC distributions with fit parameters
 are shown in Fig.~\ref{proj} with open histograms.


\bibitem{ratio}  The fraction of
$\z$ events is obtained from a one-dimensional fit to the
$M_{\mathrm{max}}(\pi J/\psi)$ distribution without including possible
interference with other amplitudes. For a more precise
determination of this fraction, a full partial wave analysis with
more statistics would be required.

\bibitem{bes3zc} M. Ablikim {\em et al.} (BESIII Collaboration), arXiv:1303.5949.


\end{thebibliography}
\end{document}